\documentclass[aps,prb,twocolumn,floatfix,showpacs]{revtex4}
\usepackage{graphicx}
\usepackage{times}
\usepackage{amsmath,amssymb}
\newcommand{\beq}{\begin{equation}}
\newcommand{\eeq}{\end{equation}}
\newcommand{\bea}{\begin{eqnarray}}
\newcommand{\eea}{\end{eqnarray}}

\newcommand{\kb}{{\bf k}}
\newcommand{\qb}{{\bf q}}

\begin{document}
\title{
Correlated Phases of 
Population Imbalanced Fermi-Fermi Mixtures on an Optical Lattice}

\author{Chen-Yen Lai}
%    \email{chen-yen.lai@email.ucr.edu} 
    \affiliation{Department of Physics and Astronomy, University of California, Riverside, California 92521, USA}
\author{Chuntai Shi}
    \affiliation{Department of Physics and Astronomy, University of California, Riverside, California 92521, USA}
\author{S.-W. Tsai}
    \email{shanwent@ucr.edu}
    \affiliation{Department of Physics and Astronomy, University of California, Riverside, California 92521, USA}

\date{\today}

%%%%%%%%%%%%%%%%%%%%%%%%%%%%%%%%%%%%%%%%%%%%%%%%%%%%%%%%%%%%%%%%%%%
\begin{abstract}
We study a two-species fermion mixture with different populations on a square lattice modeled by a Hubbard Hamiltonian 
with 
on-site interspecies repulsive interaction. Such a model can be realized in a cold atom system with fermionic atoms in two different hyperfine states loaded on an optical lattice and with tunable interspecies interaction strength via external fields.
For a two-dimensional square lattice, when at least one of the fermion species is close to half-filling, the system is highly affected by lattice effects. With the majority species near half-filling and varying densities for the minority species, we find that several correlated phases emerge as the ground state, including a spin density wave state, a charge density wave state with stripe structure, and various $p$-wave BCS pairing states for both species. We study this system using a functional renormalization group method, determine its phase diagram at weak coupling, discuss the origin and characteristics of each phase, and provide estimates for the critical temperatures.
\end{abstract}
%%%%%%%%%%%%%%%%%%%%%%%%%%%%%%%%%%%%%%%%%%%%%%%%%%%%%%%%%%%%%%%%%%%


\maketitle

%%%%%%%%%%%%%%%%%%%%%%%%%%%%%%%%%%%%%%%%%%%
%\section{Introduction}
\textit{Introduction -}
Experiments with ultra-cold atoms have realized mixtures of two different species of fermionic atoms 
with different densities,
including mixtures of 
cold lithium atoms ($^6$Li) with different populations for two different hyperfine 
states\cite{riceexp1,mitexp1,riceexp2,mitexp2,mitexp3,mitexp4}. 
Fermionic systems with imbalanced spin populations have been studied in electronic materials, such as magnetic-field-induced organic superconductors\cite{organic1,organic2}. Mixtures of different species of fermions with unequal populations have also been considered in the study of quark matter\cite{quark}. With the rapid experimental advances in the field of cold atom physics, these systems have the advantage of a great degree of tunability and control of inter-particle interactions, dimensionality, confinement, as well as number of pseudo-spin species.  
Imbalanced mixtures of cold fermionic atoms have attracted great interest due to their possible rich phase 
diagram\cite{mitexp4}. Several phases have been observed experimentally\cite{riceexp2,mitexp4}, such as imbalanced superfluid phase, phase separation, and normal Fermi liquid behavior. Also, Larkin-Ovchinnikov-Fulde-Ferrel (LOFF) state\cite{loff}, which involves Cooper pairs with finite center-of-mass momentum, and 
breached pair\cite{sarma,bpair} state with zero center-of-mass momentum, could be possible phases in special regions of the phase diagram\cite{sheehyloff}. In addition, there are studies of $p$-wave triplet pairing, caused by effective attractive intra-species interaction, proposed for strongly imbalanced cases in two\cite{kivelson2011} and in three\cite{bulgac2006,sheehy2011} dimensions.
In addition to being cooled and trapped, fermionic atoms and mixtures can now also be loaded onto optical 
lattices\cite{olexp1,olexp2} where the interaction and the hopping strengths can be tuned, and effects of their interplay with lattice geometry and dimensionality can be probed. 

%------------------------------------------------------------------------------------------------------------------------------------------------
\begin{figure}[tb]
\includegraphics[width=3.0in,angle=0]{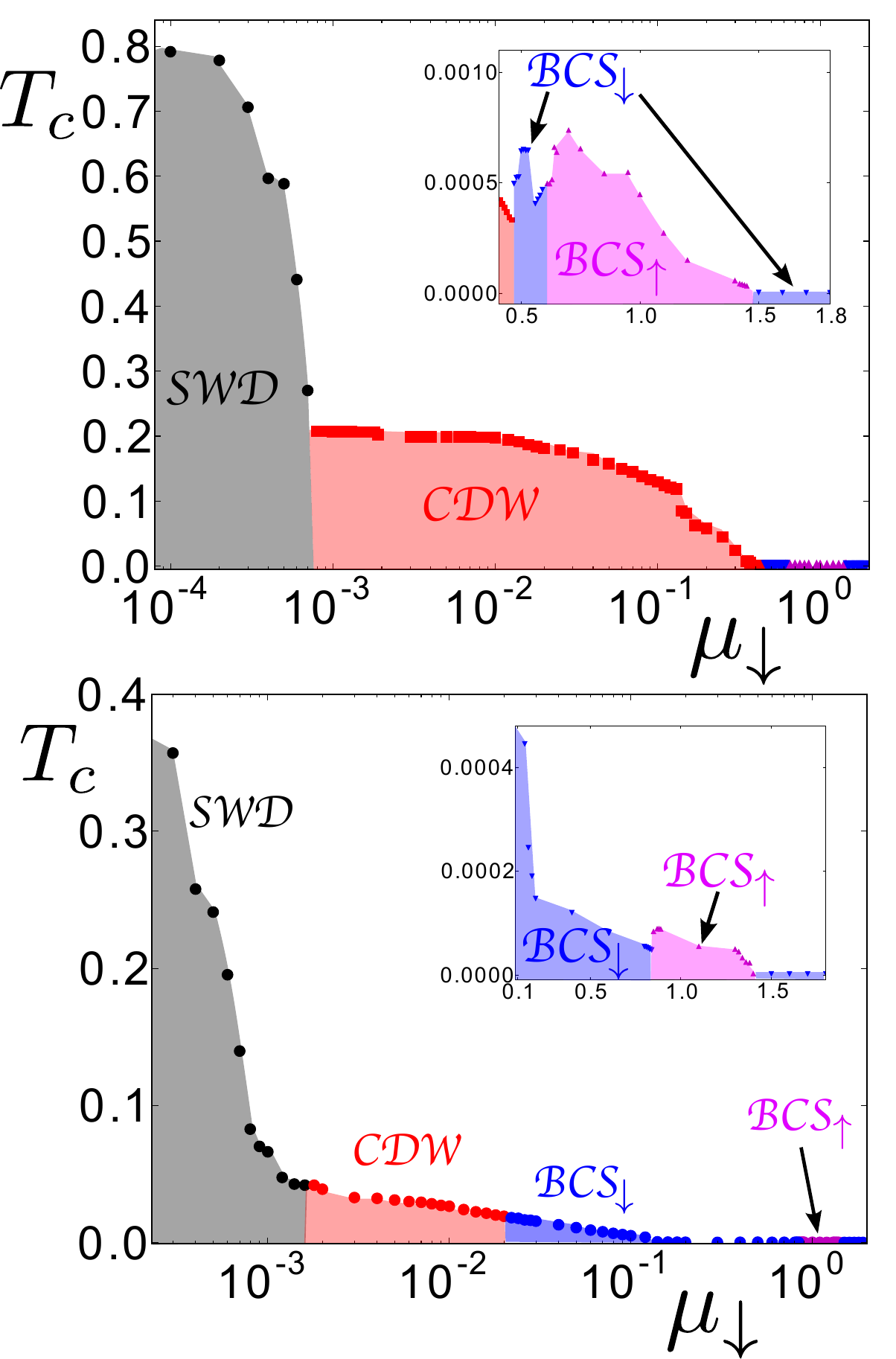}
\caption{(color on-line) Phase diagram of spin-imbalanced fermion mixture with majority species exactly at half filling $\mu_\uparrow=0$ (upper) and slightly away from half-filling $\mu_\uparrow=0.0003t$ (lower). The system goes from SDW (black) at near balance to a stripe-CDW (red) as the imbalance increases. The insert shows the strong imbalance region, where single species $p$-wave majority (magenta) or minority (blue) pairing is the dominant instability.  
}
 \label{fig:phase0}
\end{figure}
%------------------------------------------------------------------------------------------------------------------------------------------------

%In this Letter, w
We consider a two-component mixture of fermionic atoms with imbalanced populations on a two-dimensional square lattice at weak coupling region. 
It is well known that the presence of a lattice can provide interesting strong correlation effects, such as spin density wave (SDW) and charge density wave (CDW) phases for fermions on a two-dimensional (2D) square lattice at half-filling. 
Extensive studies can be found in the literature for the case of balanced, SU(2) symmetric fermions, such as the Hubbard model\cite{hubbardmodel} in various lattice geometries and 
fillings\cite{schulz1987,Shankar:1994vy,metzner2000,Zanchi:2000vv,tsai2001}.
Due to nesting of the Fermi surface (FS), SDW is the dominant instability for the repulsive Hubbard model on a 2D square lattice at half-filling, followed by $d_{x^2-y^2}$-wave superconductivity when the system is doped away from half-filling\cite{metzner2000,Zanchi:2000vv}. When nesting is completely destroyed by doping, the system becomes a normal Fermi liquid, aside from Kohn-Luttinger instabilities\cite{kl} at extremely low temperatures.
%When weak population imbalance is introduced, 
%previous studies show antiferromagnetic ordering is still dominant\cite{aforder1,aforder2,aforder3} with presence of trapping potential.
%the SDW phase is still dominant\cite{aforder1,aforder2,aforder3}. 
However, when the spin populations are unequal and $SU(2)$ symmetry is broken, both the SDW and the singlet pairing will eventually be precluded due to the mismatch of the FS for the up and down spin fermions at strong population imbalance. 
%From the $SU(2)$ symmetry argument above, singlet pairing will be allowed and dominant in attractive cases when the filling of both species are still close to each other. 
As the Fermi surfaces become increasingly mismatched, the system is expected to be dominated by other instabilities or becomes a two species Fermi liquid. If the interspecies interaction is initially repulsive, one expects to find SDW 
phase near the balanced case\cite{aforder1,aforder2,aforder3}, switching to other and potentially richer correlated behavior as the polarization increases. 
Here we study this behavior using a weak-coupling functional renormalization group (fRG) method\cite{Shankar:1994vy,metzner2000,Zanchi:2000vv,Metzner:2012jv}, which is able to treat different instabilities on an equal footing. We obtain the phase diagram for this system, focusing on the case where the majority species stays close to half-filling and the density of the minority species is varied. 
%Although only minority FS topology is changing under our consideration, this also affects the initial pattern of interspecies vertices.
The phase diagram (Fig. \ref{fig:phase0}) contains several new phases, including a stripe-CDW phase and triplet pairing phases for both species.  From our fRG study, we also obtain estimates for the critical temperature for the different instabilities.

For a microscopic interaction which is on-site, there is no bare intra-species interaction due to Pauli exclusion principle, but an effective long-range interaction can be induced via scattering between species. 
Previous studies\cite{kivelson2011,bulgac2006,sheehy2011} have considered such mediated interactions for mixtures of fermion gases (no lattice) with imbalanced populations, finding an attractive effective intra-species pairing interaction, leading to $p$-wave pairing of the majority species. In this study, we consider the effects of the interplay of interaction, population imbalance, and lattice effects. We show that lattice effects in particular, not only lead to a much richer phase diagram than that of a imbalanced mixture of fermionic gases, but also with a much higher transition temperature, even at weak interaction couplings, therefore more easily accessible to experimental observation.
Unlike 
%Such a mixture may be comparing to the 
Bose-Fermi mixtures\cite{bfmix1,bfmix2,bfmix3}, where different pairing and density wave states for the fermions originate from attractive interactions mediated by quantum fluctuations of the boson condensate, in the fermion mixture considered here, both species have screening effects from each other, and the low energy physics depends on the interplay between initial interspecies interaction, induced intra-species interaction, the FS geometry of each species, and their mismatch due to imbalance.

%%%%%%%%%%%%%%%%%%%%%%%%%%%%%%%%%%%%%%%%%%%%%%%%%%%%%%%%%%%%%%%%%%
%\section{model}
\textit{1. The model and the fRG method.}
We consider a one-band Hubbard model for each species ($\sigma=\uparrow,\downarrow$) of fermion
(with creation operator $c^\dagger_{\kb\sigma}$) 
on a 2D square lattice, with on-site 
%$s$-wave 
interspecies interaction 
$U_0$.
The Hamiltonian can be written as:
\beq
\mathcal{H}=\sum_{\sigma\kb}\xi_{\sigma\kb}c^\dagger_{\kb\sigma}c^{}_{\kb\sigma}+\frac{U_0}{V}\sum_{\kb\kb'\qb}c^\dagger_{\kb+\qb\uparrow}c^\dagger_{\kb'-\qb\downarrow}c^{}_{\kb'\downarrow}c^{}_{\kb\uparrow} \ \ 
\eeq
, where $\xi_{\sigma\kb}=-2t_\sigma(\cos k_x+\cos k_y)+\mu_\sigma$ and $V$ is the volume of the system (hereafter set to be equal to 1). The different chemical potentials $\mu_\sigma$ determine the 
densities 
and the hopping amplitude $t_\sigma$ can be tuned by the optical lattices.
In this work, we only consider the case 
$t_\uparrow=t_\downarrow=t$, and weak repulsive interspecies interaction ($U_0>0$). Previous studies\cite{melo2008} have considered the case of attractive interaction.
We neglect the confinement potential as a first approximation.

%------------------------------------------------------------------------------------------------------------------------------------------------
\begin{figure}[tb]
\includegraphics[width=1.500in,angle=270]{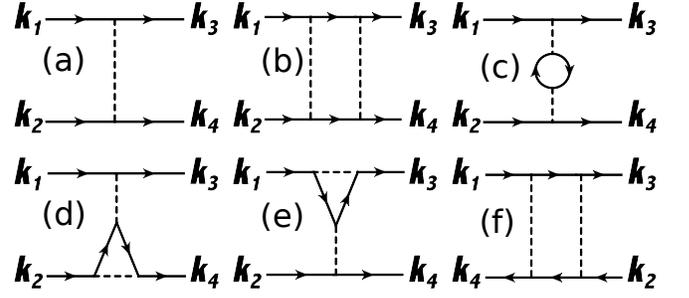}
\caption{(a) Definition of interaction vertex, where ${\bf k}_i$ is an auxiliary index including momentum, frequency, and spin. (b)-(f) The one-loop RG corrections to interaction vertices. Since initial intra-species interaction is zero, only the bubble diagram (c) can mediate intra-species vertices from 
%integrating 
two inter-species vertices.} 
\label{fig:diagram}
\end{figure}
%------------------------------------------------------------------------------------------------------------------------------------------------

%%%%%%%%%%%%%%%%%%%%%%%%%%%%%%%%%%%%%%%%%%%%%%%%%%%%%%%%%%%%%%%%%%
%\section{fRG calculation}
%\textit{2. The fRG method.}
The fRG approach has been applied to the study of the stability and instabilities of Fermi liquids\cite{Shankar:1994vy} and to a number of different lattice models\cite{Zanchi:2000vv,metzner2000,tsai2001}. 
Starting with a microscopic model, the RG method provides an effective theory for low energy scales by integrating out high energy degrees of freedom, reducing the UV cut-off $\Lambda$.
We perform a zero-temperature one-loop renormalization-group calculation for the interaction vertices $U_{\sigma\sigma^{\prime}}(\kb_1, \kb_2, \kb_3, \kb_4)$, where $\kb_1$ and $\kb_2$($\kb_3$ and $\kb_4$) are the momenta of the incoming(outgoing) fermions, and $\kb_1+\kb_2=\kb_3+\kb_4$
, shown in Fig. \ref{fig:diagram}(a).  Figs. \ref{fig:diagram}(b)-(f) show all Feynman diagrams that contribute to the renormalized four-leg vertices at one-loop. 
In order to solve the RG equations, we discretize both FS ($\sigma=\uparrow,\downarrow$) into $M$ patches and integrate the renormalization group equations numerically. We set the cut-off to be $\Lambda_l\!=\!6te^{-l}$, $dl\!=\!0.1$ for an RG step, $U_0\!=\!2.5t$ and $M\!=\!28$ in all calculations shown here. Since the bare interaction has no dynamics initially, we neglect self-energy corrections as justified for the one-loop limit at weak coupling region \cite{Shankar:1994vy,Zanchi:2000vv}. After obtaining RG flows of all marginal interaction vertices that involve states near the Fermi surfaces, we extract vertex combinations that define specific instability channels at each RG step. The pairing channels are 
$V_{BCS}^{\uparrow(\downarrow)}(\kb,\qb)=U_{\uparrow\uparrow(\downarrow\downarrow)}(\kb,-\kb,\qb,-\qb)$, and
%:
%\begin{eqnarray*}
%V_{BCS_{s,t_0}}(\kb,\qb)&=&\frac14(U_{\uparrow\downarrow}(\kb,-\kb,\qb,-\qb)\pm U_{\downarrow\uparrow}(\kb,-\kb,\qb,-\qb)\\
%&&+U_{\uparrow\downarrow}(-\kb,\kb,\qb,-\qb)\pm U_{\downarrow\uparrow}(-\kb,\kb,\qb,-\qb))
%\end{eqnarray*}
%where the upper(lower) sign stands for singlet(triplet) pairing. Density wave channels from nesting interactions are given by:
%\begin{eqnarray*}
%&&V_{SDW_z,CDW}(\kb,\qb)=\\
%&&\frac12(U_{\uparrow\uparrow}(\kb,\qb,\kb+\bf{Q},\qb+\bf{Q})+U_{\downarrow\downarrow}(\kb,\qb,\kb+\bf{Q},\qb+\bf{Q})\\
%&&\mp U_{\uparrow\downarrow}(\kb,\qb,\kb+\bf{Q},\qb+\bf{Q})\mp U_{\downarrow\uparrow}(\kb,\qb,\kb+\bf{Q},\qb+\bf{Q}))
 %,
%\end{eqnarray*}
%
$V_{BCS}^{s,t_0}(\kb,\qb)=\frac14(U_{\uparrow\downarrow}(\kb,-\kb,\qb,-\qb)\pm U_{\downarrow\uparrow}(\kb,-\kb,\qb,-\qb)
+U_{\uparrow\downarrow}(-\kb,\kb,\qb,-\qb)\pm U_{\downarrow\uparrow}(-\kb,\kb,\qb,-\qb))$,
%\end{eqnarray*}
where the upper(lower) sign stands for singlet $s$ (triplet $t_0$) pairing. Density wave channels from nesting interactions are given by
%\begin{eqnarray*}
$V_{SDW_z,CDW}(\kb,\qb)=\frac12(U_{\uparrow\uparrow}(\kb,\qb,\kb+\bf{Q},\qb+\bf{Q})+U_{\downarrow\downarrow}(\kb,\qb,\kb+\bf{Q},\qb+\bf{Q})\mp U_{\uparrow\downarrow}(\kb,\qb,\kb+\bf{Q},\qb+\bf{Q})\mp U_{\downarrow\uparrow}(\kb,\qb,\kb+\bf{Q},\qb+\bf{Q}))$
where the upper(lower) sign stands for SDW$_z$(CDW) channel, and $\bf{Q}=(\pm\pi,\pm\pi)$ is a nesting vector.
The pseudo-spin index $\sigma$ refers to two different hyperfine states of the atoms, thus the term SDW is understood as ordering of the pseudo-spin.
%, and not of real spin. 
Also, the cold atoms are neutral particles and CDW refers to the number density (not electric charge), but we keep the CDW terminology in analogy to electronic systems.
At each RG step, we diagonalize these $M\times M$ matrices as $\sum V_{a}(\kb,
\qb)\phi_a^{(n)}(\qb)=\lambda_a^{(n)}\phi_a^{(n)}(\kb)$. As the leading eigenvalue of a defined channel diverge, we identify it as an instability of the system (the BCS pairing instability requires a negative eigenvalue). 
The form of the corresponding eigenvector determines the symmetry of the order parameter of the instability. 
The scale $\Lambda_c$ at which the divergence occurs can be associated to a mean-field critical temperature\cite{Zanchi:2000vv}.
%, as $\Lambda_c\sim T_c$. 

%------------------------------------------------------------------------------------------------------------------------------------------------
\begin{figure}[tb]
\includegraphics[width=3.300in,angle=0]{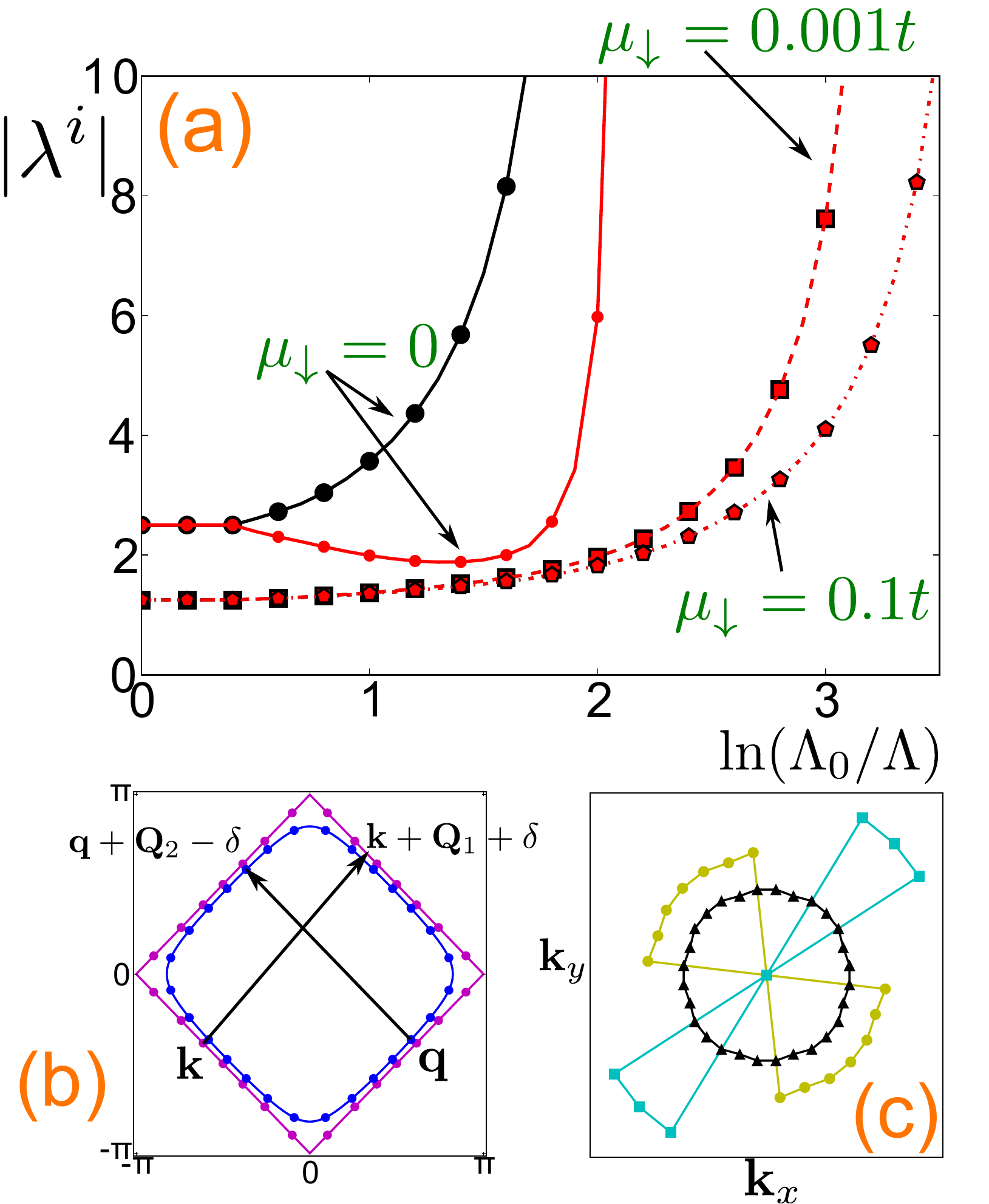}
\caption{
(a) RG flow of $SDW_z$ (black) and $CDW$ (red) channels 
at different minority fillings, where $\mu_\downarrow=0$ (solid), $\mu_\downarrow=0.001t$ (dash), and $\mu_\downarrow=0.1t$ (dots) with majority species half-filled ($\mu_\uparrow=0$). (b) One example of non-perfect nesting particle-hole process with net momentum $(2\pi,0)$. (c) The symmetry of the order parameter of majority CDW (yellow circle) and minority CDW (cyan square) for the stripe-CDW phase ($\mu=0.01t$). A conventional $s$-wave form factor (black triangular) is also shown. Both species behave like stripe density wave along the diagonal direction. The circles on the FS are fRG patches.}
\label{fig:rgflowdw}
\end{figure}
%------------------------------------------------------------------------------------------------------------------------------------------------

%%%%%%%%%%%%%%%%%%%%%%%%%%%%%%%%%%%%%%%%%%%%%%%%%%%%%%%%%%%%%%%%%%
%\section{Results}
\textit{2. Nearly balanced.}
In order to maximize lattice effects, we focus on the case where the majority species is at or close to half-filling, $\mu_\uparrow\!\sim\!0$, and we vary 
$\mu_\downarrow$. Phase diagrams are shown in Fig. \ref{fig:phase0}.
When the system is balanced ($\mu_\uparrow\!=\!\mu_\downarrow\!=\!0$), SDW is the dominant instability (RG flows shown in Fig. \ref{fig:rgflowdw}(a)). With increasing $\mu_{\downarrow}$, 
the SDW$_z$ order persists for the nearly balanced region (Fig. \ref{fig:phase0}), but the critical temperature $T_c$ decreases. This result agrees with previous studies\cite{aforder1} showing that antiferromagnetic order is suppressed by increasing imbalance (i.e. $\Delta\mu$).
%Fig.\ref{fig:rgflow0}(c) and (d) show  intensity plots for $U_{\uparrow\uparrow}(k_1,k_2,k_3)$ and $U_{\uparrow\downarrow}(k_1,k_2,k_3)$, at a given RG step, as $k_2$ and $k_3$ are varied, covering all patches around the FS (patch index $n$ from 0 to 27), with $k_1$ fixed at patch $n_1=1$. 
%There is an enhanced vertical line at $n_3 \!=\! 19$ in these two figures since these patches have $k_1 \!=\! k_3\! +\! (\pi, \pi)$, that is, they are related by a nesting vector, and enhanced diagonal lines corresponding to nesting between $k_2$ and $k_3$ in Fig.1(d).

\textit{3. Weak imbalance.}
As the minority density is further decreased away from half-filling, some interspecies vertices involving nesting vectors, such as \textit{umklapp} processes, are no longer near the FS and are thus suppressed. 
%Due to the suppressed interspecies nesting, the antiferromagnetic order decreases and other instability dominates.
%Since the majority species are still perfectly nested, and partial interspecies nesting vertices are still allowed.
%
However, there are still some non-perfect nesting particle-hole processes with net momentum equal to a reciprocal lattice vector, $(2\pi,0)$ or $(0,2\pi)$, which are allowed even under imbalance. 
An example of such a process, depicted in Fig. \ref{fig:rgflowdw}(b), is for a minority fermion to scatter from ${\bf q} \rightarrow {\bf q} + {\bf Q}_2 - \boldsymbol{\delta}$ across two opposite sides of its FS (say along the $45^{\circ}$ direction), while a majority fermion scatters from ${\bf k} \rightarrow {\bf k} + {\bf Q}_1 + \boldsymbol{\delta}$ across the other two opposite sides of its FS (say along the $135^{\circ}$ direction). With ${\bf Q}_1 = (\pi, \pi)$, ${\bf Q}_2 = (-\pi, \pi)$, and $\boldsymbol{\delta} = (\delta, \delta)$ accounting for the FS mismatch, the net momentum for this process is $(0, 2\pi)$ and therefore of {\it umklapp}-type and allowed by momentum conservation in spite of the imbalance.
These vertices can still renormalize significantly under RG flow, and more importantly, they mediate intra-species nesting processes for the majority fermions, which do have a perfectly nested FS, so these mediated processes have the strongest flow. On the other hand, the induced intra-species long-range attraction is maximum at diagonal directions. Due to the partial nesting and intra-species long-range attraction, this leads to a crossover in the RG flows from the SDW$_z$ to a stripe-CDW phase.
From a real space picture, with increasing imbalance, both species also have more empty sites to move around to minimize free energy. 
Fig. \ref{fig:rgflowdw}(a) shows the RG flow close to this crossover, where SDW$_z$ and CDW channels are almost degenerate. 
%The lower panel of Fig.\ref{fig:rgflow0}(b) shows the corresponding order parameter symmetries. 
%At the mean time, t
The CDW channel of both species is doubly degenerate. 
%For one of them, shown in 
The order parameters for $\uparrow$ and $\downarrow$ are non-zero on two-opposite sides of the FS (second and fourth quadrants), and zero on the other sides (first and third quadrants), with the situation reversed for the other degenerate channel (Fig. \ref{fig:rgflowdw}(c) shows one of the cases). These correspond to stripe charge order in the diagonal direction\cite{Bhongale:2012fb}. 
This stripe-CDW phase originates from weaker spin fluctuations caused by imbalance in addition to screened intra-species long-range attraction.
%Figs.\ref{fig:rgflow}(a) shows snapshots of the interspecies and majority intra-species vertices as the instability is approached. 
%Comparing with the $SU(2)$ symmetric case (Fig.\ref{fig:rgflow0}(c) and \ref{fig:rgflow0}(d)), where the interspecies nesting interaction is the dominant one, when the imbalance is increased ($\mu_\downarrow\!=\!0.01t$), the inter and intra-species nesting vertices are comparable and carry different signs, indicating that this CDW stripe phase originates from weaker spin fluctuations caused by imbalance in addition to screened intrapecies attraction.
Since it involves both spin and charge fluctuations, the size of this region and its critical energy scale in phase diagram depend on the filling fraction of both species, shown in Fig. \ref{fig:phase0}.

%------------------------------------------------------------------------------------------------------------------------------------------------
\begin{figure}[tb]
\includegraphics[width=3.300in,angle=0]{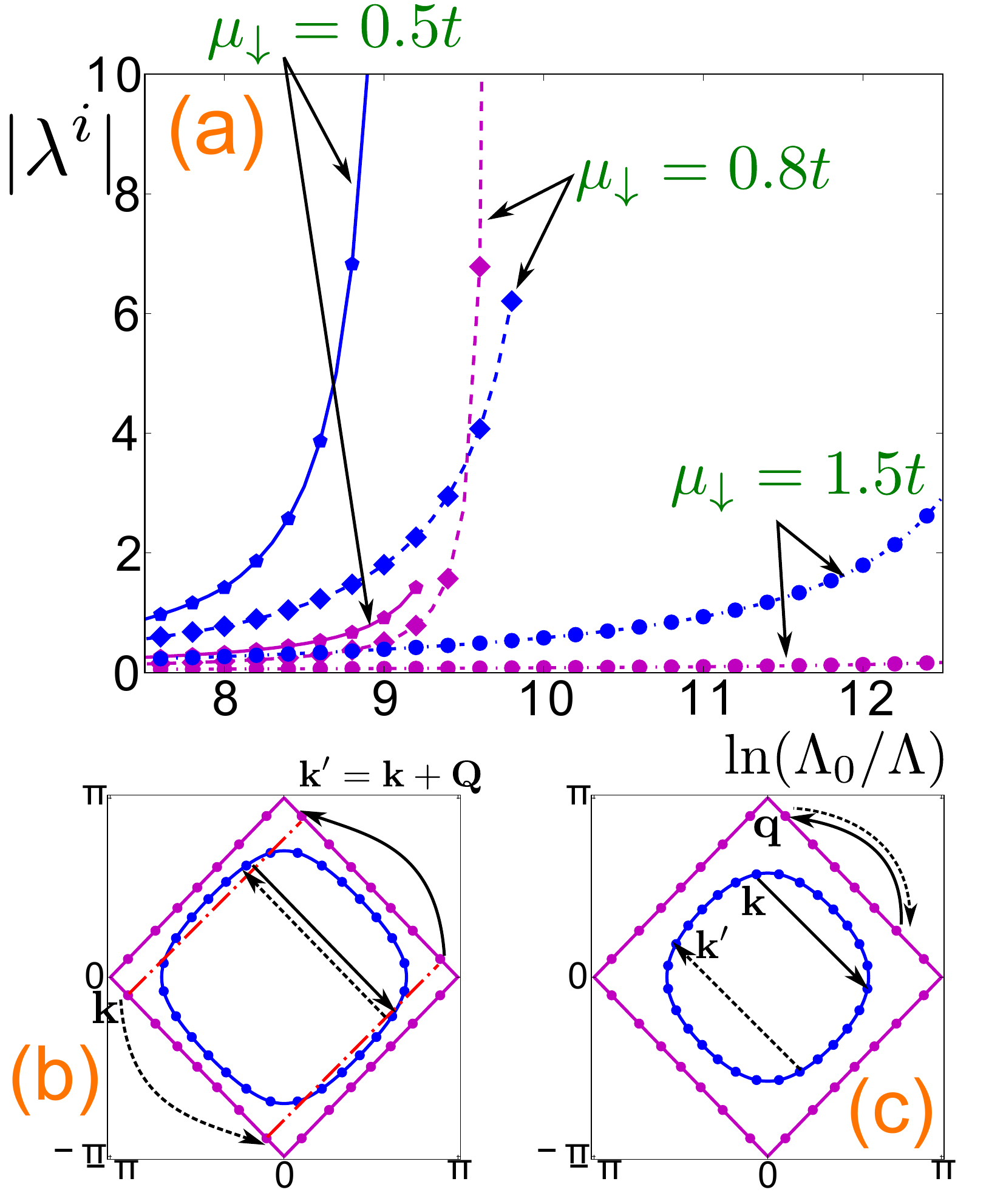}
\caption{
(a) RG flow of majority (blue) and minority (magenta) $p$-wave BCS channels
at three different minority fillings, where $\mu_\downarrow=0.5t$ (solid), $\mu_\downarrow=0.8t$ (dash), and $\mu_\downarrow=1.5t$ (dot) with majority species half-filled ($\mu_\uparrow=0$). In (b), $\mu_\downarrow=0.8t$ (blue), the renormalized majority pairing channel $U_{\uparrow\uparrow}(\kb,-\kb,-\kb^{\prime},\kb^{\prime})$ couples to its own nesting channel $U_{\uparrow\uparrow}(\kb,-\kb,-\kb+{\bf Q},\kb+{\bf Q})$. In (c), $\mu_\downarrow=1.5t$, the half-filled majority species can provide more phase space to renormalize minority pairing channel (i.e. $q$ can be anywhere on that FS branch).}
\label{fig:rgflowbcs}
\end{figure}
%------------------------------------------------------------------------------------------------------------------------------------------------
\textit{4. Strong imbalance.}
As the imbalance becomes stronger, the FS mismatch precludes all nesting processes as well as zero-momentum singlet pairing. Since the induced intra-species interactions are always attractive, the dominant RG flows present another crossover, from CDW stripe phase to triplet $p$-wave BCS pairing. The triplet pairing is for a single fermion species but it is generated by the initial bare repulsive interspecies interaction, the lattice (FS shape), and the imbalance (FS mismatch). Whether the dominant pairing is for the majority or for the minority species depends on the interplay of these factors.
Fig. \ref{fig:rgflowbcs}(a) shows RG flows for three different minority chemical potentials. As $\mu_{\downarrow}$ 
increases, 
the minority pairing channel is dominant at first. This is because the nearly half-filled majority FS provides a large phase space in RG process to mediate and renormalize the minority intra-species BCS vertex. 
The bubble diagram in Fig. \ref{fig:diagram}(c), 
which contains an internal fermion loop, gives a first nonzero correction to the intra-species vertex from two interspecies vertices, of the form,
\beq
\partial_l U_{\sigma\sigma'}^{(l)}(\kb_1,\kb_2,\kb_3)\!=\!\left[1\!-\!\frac12(1\!+\!X)\delta_{\sigma\sigma'}\right]\!\!
\beta_{ph}\{U_1,U_2\},
\eeq
where $X$ denotes the operation $XF(1,2,3,4)=F(2,1,3,4)$
and $\beta_{ph}\{U_1,U_2\}=\Pi\{U_1,U_2\}+\mathcal{T}\Pi\{U_1,U_2\}$, with $\mathcal{T}F(1,2,3,4)=F(3,4,1,2)$ the time-reversal operator and 
\bea\label{bph1}
\Pi\{U_1,\!U_2\!\}
\!=\!\!\!\sum_{{\bf q},\alpha}\!\!B_{ph}\!(\kb_1,\!\kb_3,\!{\bf q})U_{\!\sigma\!\alpha}\!(\kb_1,\!{\bf q},\!\kb_3)U_{\!\sigma'\!\alpha}\!(\kb_4,\!\qb,\!\kb_2)
\eea
\noindent where $B_{ph}$ is an integral over an angular sector
\cite{Zanchi:2000vv}. 
From Eq. (\ref{bph1}), the induced intra-species BCS vertex comes from integrating out two interspecies interactions. 
For example, in Fig. \ref{fig:rgflowbcs}(b), integrating pairs of solid and dash lines, which represent two interspecies vertices, results in an intra-species BCS vertex, $U_{\downarrow\downarrow}(\kb,-\kb,\kb^{\prime},-\kb^{\prime})$.
Because the FS of the majority species is flat, there will be more renormalization corrections to intra-species interaction between minority species at first. 
As shown in Fig. \ref{fig:rgflowbcs}(c), the momentum $\qb$ at fixed magnitude can be translated anywhere along that branch of the FS. 
Fig. \ref{fig:phase0} shows that the critical temperature(a.u.) for minority pairing decreases monotonically with increasing polarization, because the phase space of induced minority intra-species pairing becomes smaller. 
The same effect is observed for the majority $p$-wave pairing channel, which is the sub-leading channel.
%This makes the majority pairing become the dominant instability.
However, the induced BCS vertex of majority species couples to its own nesting channel ($\kb$ and $\kb'$, for example,  are connected by a nesting vector in Fig. \ref{fig:rgflowbcs}(b)). Although the initial mediated majority intra-species pairing interaction is smaller 
%(due to phase space) 
than that for minority species, as discussed above, eventually the strong RG flow through a nesting channel leads to a majority pairing instability. We note that this majority superfluid is not the same as proposed in previous studies of fermion gases\cite{kivelson2011,bulgac2006,sheehy2011}, but is instead a lattice effect and has a much larger energy scale and critical temperature. 
As the density of minority species is further decreased, 
it starts to behave like 
a fermion gas, 
with a small, almost circular FS, as shown for example in Fig. \ref{fig:rgflowbcs}(c). 
When the density of minority species becomes less than quarter filling, 
the induced majority BCS vertices are no longer coupled to its own nesting vertices, which causes a 
%non-monotonically 
sudden drop of the critical temperature for the majority pairing channel. 
Since the majority species is still perfectly nested
%, as discussed earlier, 
there is more phase space for the minority pairing interactions to be mediated and renormalized. 
This crossover of RG flows is shown in Fig. \ref{fig:rgflowbcs}(a). The majority pairing instability only exists in a small region of the phase diagram for each different majority filling, as shown in Fig. \ref{fig:phase0}. 
%majority BCS vertices involving fermions on opposite sides of the nested FS can no longer be mediated by minority fermions near the minority FS since this momentum would not be conserved. 
For large minority chemical potentials, the critical temperature is much smaller, and eventually reaches our numerical limit ($\Lambda_l\!\sim\!10^{-5}t$). 
Before this limit is reached, the flow of minority pairing is stronger than majority pairing and other channels. 
%Although we do not obtain a divergence in the minority pairing channel, the system is expected to have a Kohn-Luttinger instability of minority pairing in a rather low critical temperature.
% based on the phase space argument above.

%%%%%%%%%%%%%%%%%%%%%%%%%%%%%%%%%%%%%%%%%%%%%%%%%%%%%%%%%%%%%%%%%%
%\begin{figure}[tb]
%\includegraphics[width=2.60in,angle=270]{phase3.pdf}
%\caption{Phase diagram of spin-imbalanced fermion mixture when majority species slightly away from half filling ($\mu_\uparrow=0.003t$). } 
%%(b)-(c) The estimated critical temperature, $\Lambda_c\!=\!6te^{-l_c}$, of each dominate phase of repulsive model at different majority fillings.}
%\label{fig:phase}
%\end{figure}
%%%%%%%%%%%%%%%%%%%%%%%%%%%%%%%%%%%%%%%%%%%%%%%%%%%%%%%%%%%%%%%%%%
  
%%%%%%%%%%%%%%%%%%%%%%%%%%%%%%%%%%%%%%%%%%%
%\section{Conclusion}
\textit{5. Conclusion.}
We have performed a weak-coupling fRG study of a population imbalanced fermion system on a square optical lattice in two dimensions. 
%When the majority species is close to half-filing, both density wave and triplet pairing phases are highly favorable due to the nesting of majority species FS. 
At the weak imbalance region, the competition between spin fluctuations and intra-species attractions leads to a stripe density wave phase, rather than the usual uniform SDW phase.
As the imbalance is further increased, $p$-wave superfluid phases become dominant due to the mismatch of the FS. 
Although the $p$-wave pairing superfluid may be expected in dilute density limit, 
the competition between majority and minority pairing is determined by their FS topology and the mediated intra-species pairing interaction which is renormalized from the initial bare interspecies on-site repulsive interaction.
Both the stripe density wave phase and triplet superfluid phases are enhanced by nesting of the majority FS, leading to much higher critical temperatures than that found for the imbalanced Fermi gas without  lattice\cite{kivelson2011,bulgac2006} at weak coupling limit. 
%Our result shows presence of lattice and population imbalance are important for this rich phase diagram and higher critical temperature,  which is different from Bose-Fermi mixture\cite{bfmix1,bfmix2} and imbalanced fermi gas without lattice\cite{kivelson2011,bulgac2006} at weak coupling limit. 
According to our calculation, the critical temperature for stripe density wave is roughly $\Lambda^{CDW}_c(\mu_\uparrow\!=\!0,\mu_\downarrow\!=\!0.01t)\!\approx\!0.18t$ and that for $p$-wave superfluid is $\Lambda^{BCS_\downarrow}_c(\mu_\uparrow\!=\!0.0003t,\mu_\downarrow\!=\!0.03t)\!\approx\!0.016t$, where the hopping amplitude $t$ can be controlled experimentally. 
In a recent experiment\cite{riceexp}, it is reported that temperatures around 6\% of the Fermi temperature of a noninteracting trapped gas has been achieved.
The symmetry of the order parameter of each instability can be probed by momentum-resolved spectroscopy\cite{rfexp1,rfexp2}.
These techniques provide analogues of angle-resolved photoemission spectroscopy (ARPES) used in condensed matter systems, and can probe anisotropic systems, such as the stripe CDW and the $p$-wave pairing that we predict here.

%%%%%%%%%%%%%%%%%%%%%%%%%%%%%%%%%%%%%%%%%%%%%%%%%%%%%%%%%%%%%%
%\begin{Acknowledgments}
\textit{Acknowledgments - }
We thank Pochung Chen for the computational resources in National Tsing Hua University, Taiwan. SWT thanks Tun Wang for useful discussions in the early stages of this work. We gratefully acknowledge support from NSF under Grant DMR-0847801 and from the UC-Lab FRP under Award number 09-LR-05-118602.
%\end{acknowledgments}
%%%%%%%%%%%%%%%%%%%%%%%%%%%%%%%%%%%%%%%%%%%%%%%%%%%%%%%%%%%%%%

\end{document}